\documentclass[traditabstract]{aa}  
\usepackage{graphicx}
\usepackage{txfonts}
\usepackage{hyperref}
\usepackage{rotating}
\usepackage{natbib}
\usepackage[usenames]{color}
\usepackage{multirow}
\usepackage{url}
\usepackage{ulem}
\bibpunct{(}{)}{;}{a}{}{,} 

\newcommand{\kms}   {{\rm \  km \  s^{-1}}}

\begin{document}

\title{ Dense gas in low-metallicity galaxies }
		
\author{J. Braine\inst{1} \and Y. Shimajiri\inst{2}  \and P. Andr\'e\inst{2} \and S. Bontemps\inst{1} \and Yu Gao\inst{3,4} \and Hao Chen\inst{5,6,7} \and C. Kramer\inst{8} }

  \institute{Laboratoire d'Astrophysique de Bordeaux, Univ. Bordeaux, CNRS, B18N, allŽe Geoffroy Saint-Hilaire, 33615 Pessac, France.\\
             \email{jonathan.braine@u-bordeaux.fr}
        \and
            Laboratoire AIM, CEA/DRF--CNRS--Universit\'e Paris Diderot, 
            IRFU/Service dÕAstrophysique, C.E. Saclay, Orme des Merisiers, 91191 Gif-sur-Yvette, France
        \and
             Purple Mountain Observatory, Chinese Academy of Sciences, 2 West Beijing Road, Nanjing 210008, P. R. China;
        \and
             Key Laboratory of Radio Astronomy, Chinese Academy of Sciences, Nanjing 210008, P. R. China
        \and               
	   School of Astronomy and Space Science, Nanjing University, Nanjing 210093, P. R. China
        \and
               Key Laboratory of Modern Astronomy and Astrophysics, Nanjing University, Nanjing 210093, P. R. China
        \and
           Collaborative Innovation Center of Modern Astronomy and Space Exploration, Nanjing 210093, P. R. China
        \and
             Instituto Radioastronom'a Milim\'etrica, Av. Divina Pastora 7, Nucleo Central, 18012 Granada, Spain
  }
\date{Received xxxx; accepted xxxx}

\abstract {Stars form out of the densest parts of molecular clouds. Far-IR emission can be used to estimate the Star Formation Rate (SFR) and high dipole moment molecules, typically HCN, trace the dense gas. A strong correlation exists between HCN and Far-IR emission, with the ratio being nearly constant, over a large range of physical scales. A few recent observations have found HCN to be weak with respect to the Far-IR and CO in subsolar metallicity (low-Z) objects.  We present observations of the Local Group galaxies M~33, IC~10, and NGC~6822 with the IRAM 30meter and NRO 45m telescopes, greatly improving the sample of low-Z galaxies observed.  HCN, HCO$^+$, CS, C$_2$H, and HNC have been detected. Compared to solar metallicity galaxies, the Nitrogen-bearing species are weak (HCN, HNC) or not detected (CN, HNCO, N$_2$H$^+$) relative to Far-IR or CO emission. HCO$^+$ and C$_2$H 
emission is normal with respect to CO and Far-IR. While $^{13}$CO is the usual factor 10 weaker than $^{12}$CO, C$^{18}$O emission was not detected down to very low levels. Including earlier data, we find that the HCN/HCO$^+$ ratio varies with metallicity (O/H) and attribute this to the sharply decreasing Nitrogen abundance. The dense gas fraction, traced by the HCN/CO and HCO$^+$/CO ratios, follows the SFR but in the low-Z objects the HCO$^+$ is much easier to measure.  Combined with larger and smaller scale measurements, the HCO$^+$ line appears to be an excellent tracer of dense gas and varies linearly with the SFR for both low and high metallicities. }	


\maketitle

\section{Introduction}


Stars are believed to condense out of prestellar cores within the densest 
parts, clumps and/or filaments above $A_v \sim 8$ or $n_{H2} \sim 1-2 \times 10^4$ cm$^{-3}$ 
\citep{Andre2010,Lada2010} of giant molecular clouds (GMCs).  
To study the link between gas and star formation, it is important to observe molecules which emit
preferentially in these environments.  HCN and HCO$^+$ are used to trace dense molecular gas
because their dipole moments are high, such that they are thermalized by collisions only at high 
densities (n(H$_2$) $> 10^4$cm$^{-3}$) even for the lowest transitions.  If star formation proceeds 
in a quasi-universal manner in all dense clumps/filaments \citep{Andre14}, then we expect the star 
formation rate (SFR) to be strongly and linearly correlated with the dense gas mass.  In terms of observables, we expect 
the Far-InfraRed (FIR) emission to follow that of HCN and HCO$^+$.  A large number of articles  
 \citep[e.g.][]{Gao04b,Gao04,Wu2005,Gao07, Baan08,Gracia-Carpio08,Wu2010,Garcia-Burillo2012,Liu2010,Lada2012,Kepley2014}  have  studied the link between
HCN, and HCO$^+$ in more recent work, with far-infrared (FIR) luminosities in a variety of 
environments.  They tend to show that FIR and HCN emission (and often HCO$^+$) are closely linked.
Recent work by \citet{Chen2015} and \citet{Usero2015} showed that HCN emission, both with respect to the total molecular gas mass (as traced by CO) and to the SFR (as traced by Far-IR emission), was significantly stronger in galactic nuclei as compared to galactic disks.  

Both HCN and HCO$^+$ have low abundances, orders of magnitude lower than CO, so their abundances can 
be significantly modified by changes in the chemical network leading to their formation \citep[e.g. ][]{Lopez-Sepulcre10}, and these changes are not necessarily be linked to star formation.  Unlike HCN, HCO$^+$ is a molecular ion and affected by the ionization equilibrium.  In addition, the collisional cross-section of HCO$^+$ is much larger than that of HCN.  Both molecules are linear and thus are described by a single rotational quantum number.  Elemental abundance ratios are not always constant and can affect the production of molecules.


In particular, the Nitrogen abundance tends to decrease more quickly than the Oxygen or Carbon in low-metallicity galaxies.  In galactic disks, the HCN/CO and HCO$^+$/CO line intensity ratios (1--0 transition, expressed in Kkm/s) are about 2\%, although in galactic nuclei and starburst galaxies these ratios tend to be higher (see references in first paragraph and \citet{Brouillet05} for M~31 and \citet{Kuno95} for M~51).  Recently, \citet{Buchbender13} showed that in the roughly half-solar metallicity galaxy M~33, the emission from the dense gas tracers HCN and HCO$^+$ was weak relative to CO, particularly HCN.  \citet{Gratier10b} actually failed to detect HCN in the CO-bright region Hubble V in the Local Group low-metallicity galaxy NGC 6822, with a $1\sigma$ limit 180 times weaker than the CO.  The variations between HCO$^+$ and HCN show that new observing programs require both tracers as HCO$^+$ is probably less affected by metallicity variations and tends to be stronger than HCN in disks but not in nuclei.

Descending the mass/morphology/metallicity sequence from M~31 to M~33 to the Magellanic Clouds (MC), the HCO$^+$/HCN ratio increases from 1.2 in M~31 \citep[][particularly figure 3]{Brouillet05}, 1.5 (range from 1.1 to 2.5) in M~33 \citep{Buchbender13}, to $\sim 1.8$ in the Large MC \citep{Chin97} and 3 in the Small MC \citep[][single position]{Chin98}.  Within M~31, a rather rough trend can be seen with radius, with the HCO$^+$/HCN ratio increasing from 1 in the inner parts to $\sim 1.4$ in the outer disk.  
Are these variations in the HCN/CO ratio and the HCO$^+$/HCN ratio due to metallicity?  

Recent comparisons between Galactic and extragalactic studies of the SF rate \citep{Heiderman10, Lada2012} suggest that there may a quasi-universal "star 
formation law" such that dense gas ($\ge 10^4$ cm$^{-3}$) is converted into stars at a constant rate with a constant stellar initial mass function (IMF). It has 
been proposed that this may be the result of the quasi-universal filamentary structure 
of molecular clouds \citep{Andre14}.

In this work, our goal is to understand how to use dense gas tracers in low-metallicity galaxies.  More precisely, we wish to measure the variation with metallicity of the link between the dense gas tracers HCN and HCO$^+$ and the FIR and CO emission.  In order to keep observing times reasonable, only positions/objects with CO lines brighter than $\sim$0.5K (T$_{mb}$ at $\sim20"$ resolution) were observed.  In order that the linear resolution be comparable to the size of a molecular cloud or cloud complex ($\sim 100$pc), 
we restricted our observations to Local Group galaxies which cover a rather complete range in metallicity.
Most extragalactic studies focus on the central regions -- whether by choice or simply due to the single pointing per object -- where it is known that gas is on average denser.  By observing Local Group objects, we observe identifiable positions in disks where a wealth of other data is available, such as HI column densities and several tracers of the SFR (H$\alpha$, FIR, FUV...).  
Detailed analyses of the biases and validity of a broad variety of SFR tracers can be found in \citet{Boquien11b} and \citet{Galametz2013}.

While the major starbursts (ULIRGs) observed in \citet[e.g.][]{Gao04,Gao04b} likely have a solar metallicity, this is not the general case for distant galaxies.  Only by studying the role of metallicity in local galaxies can we evaluate the "universality" of the "dense gas -- star formation" relation and its tentative application to high-redshift galaxies.  

In addition to HCN(1--0) and HCO$^+$(1--0), the HNC(1--0), N$_2$H$^+$(1--0), CS(2--1), and CCH lines were observed and, with the exception of N$_2$H$^+$, detected.  The HCN/HNC ratio is typically about 3 in the Magellanic clouds \citep{Chin97} but closer to unity in e.g. IC342 \citep{Meier05}.  Galactic observations suggest that the HCN/HNC ratio can be used as a probe of physical conditions (HCN and HNC dipole moments are virtually identical). 
The HCN/HNC ratio is typically greater than unity, sometimes much greater, in warm GMC cores \citep[e.g.][]{Hirota98} but can be as low as 0.1 or 0.2 in dark cloud cores \citep[e.g. ][ and references therein]{Tennekes06}.  Thus, detected variations in this ratio are likely a sensitive temperature indicator.
The N$_2$H$^+$ line is a classical tracer of cool and dense regions in the Galaxy as it depletes only very weakly onto dust grains.  With 2 Nitrogen atoms, N$_2$H$^+$ should be strongly affected by metallicity variations, providing a check for the HCN/HCO$^+$ ratio as a function of metallicity. 

\begin{figure}
	\centering
	\includegraphics[width=\hsize{}]{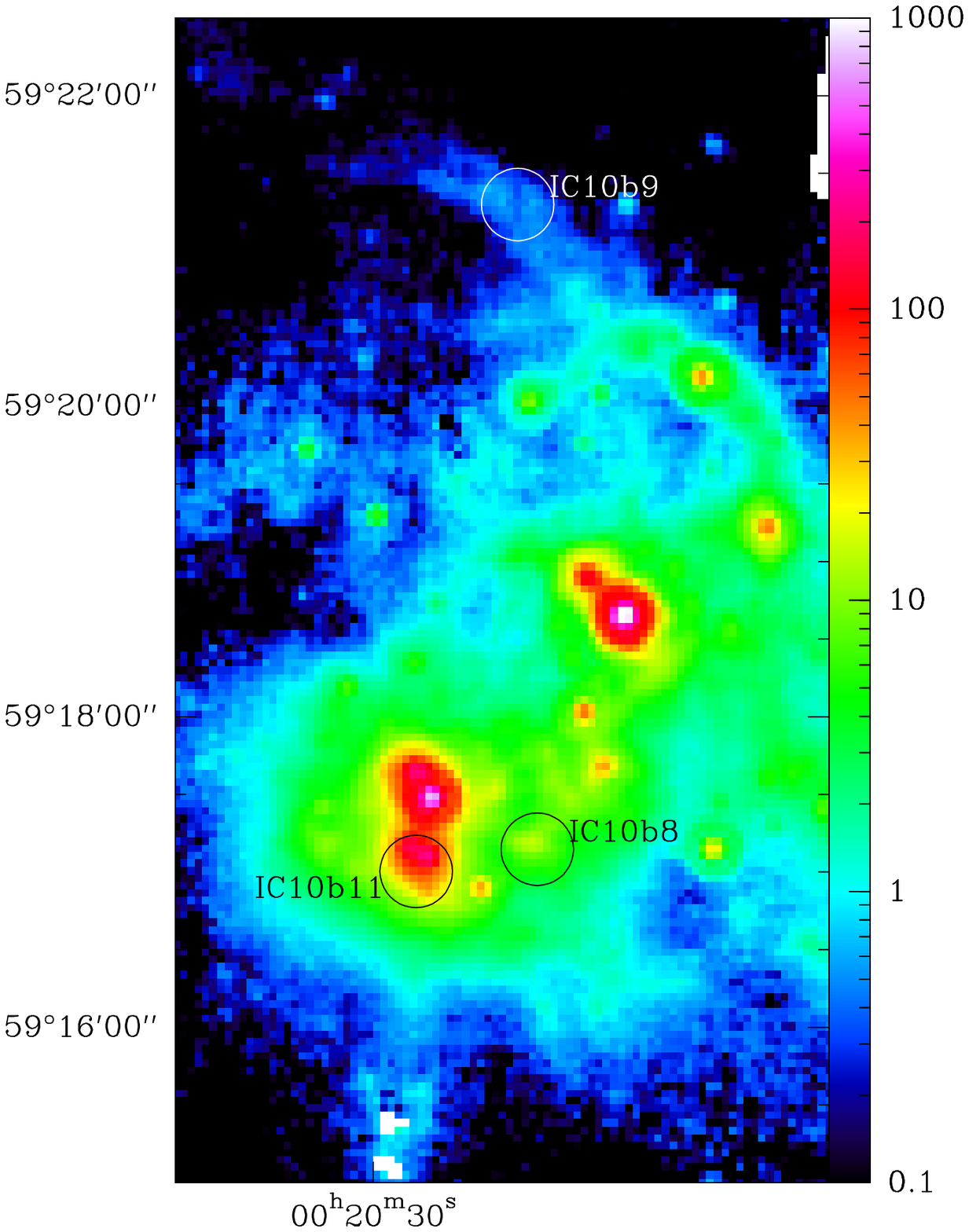}
	\caption{Observed positions on a Spitzer 24$\mu$m map of IC10.  The circle size indicates the beamwidth of the observations.  Units of the 24$\mu$m emission are MJy/sr and the color scale is the same for Figures 1--3 in order to facilitate comparison. }
	\label{ic10_24mic} 
\end{figure}

\begin{figure}
	\centering
	\includegraphics[width=\hsize{}]{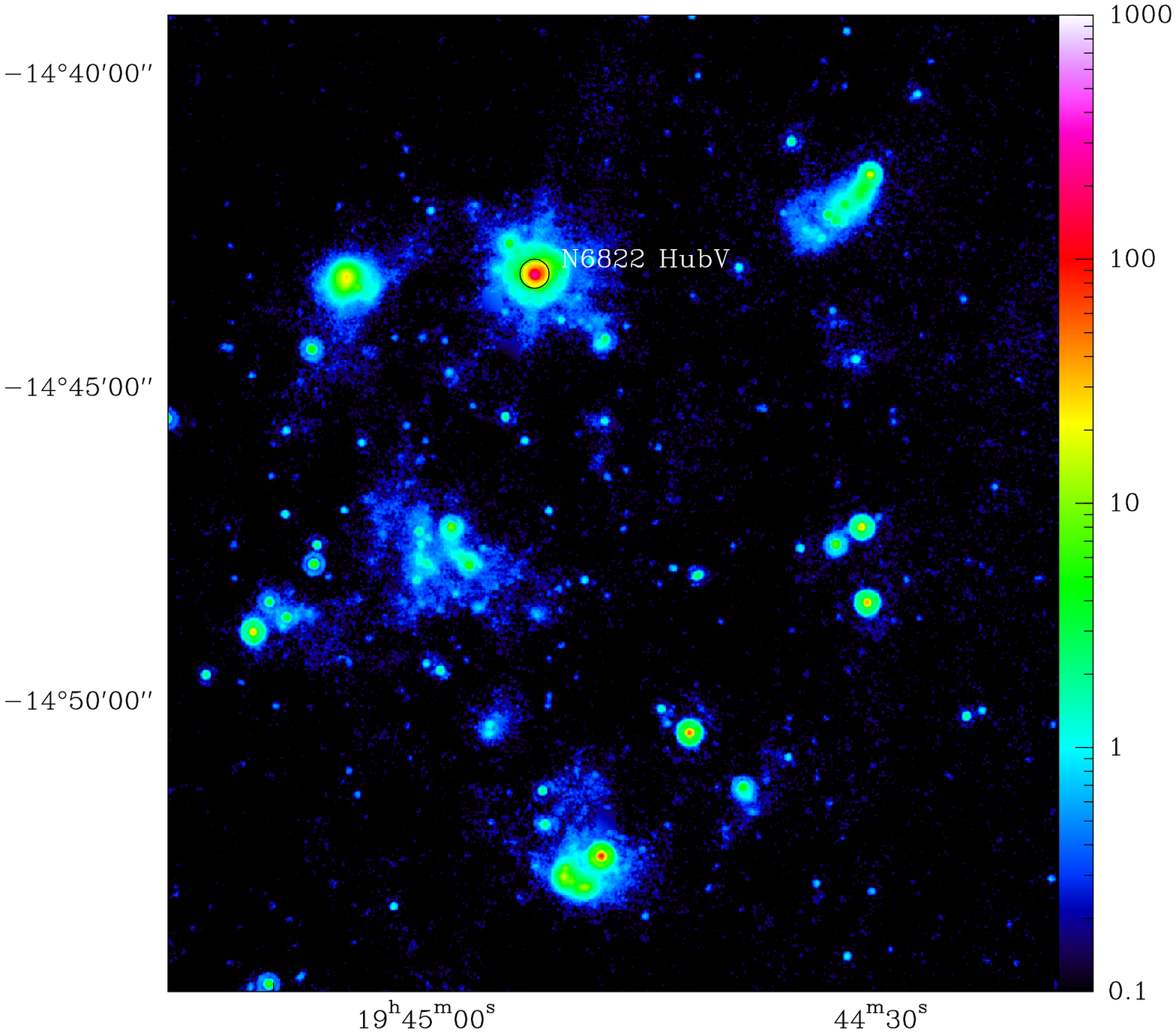}
	\caption{Observed position on a Spitzer 24$\mu$m map of NGC~6822.  The circle size indicates the beamwidth of the observations.  Units of the 24$\mu$m emission are MJy/sr and the color scale is the same for Figures 1--3 in order to facilitate comparison.}
	\label{n6822_24mic} 
\end{figure}

\begin{figure}
	\centering
	\includegraphics[width=\hsize{}]{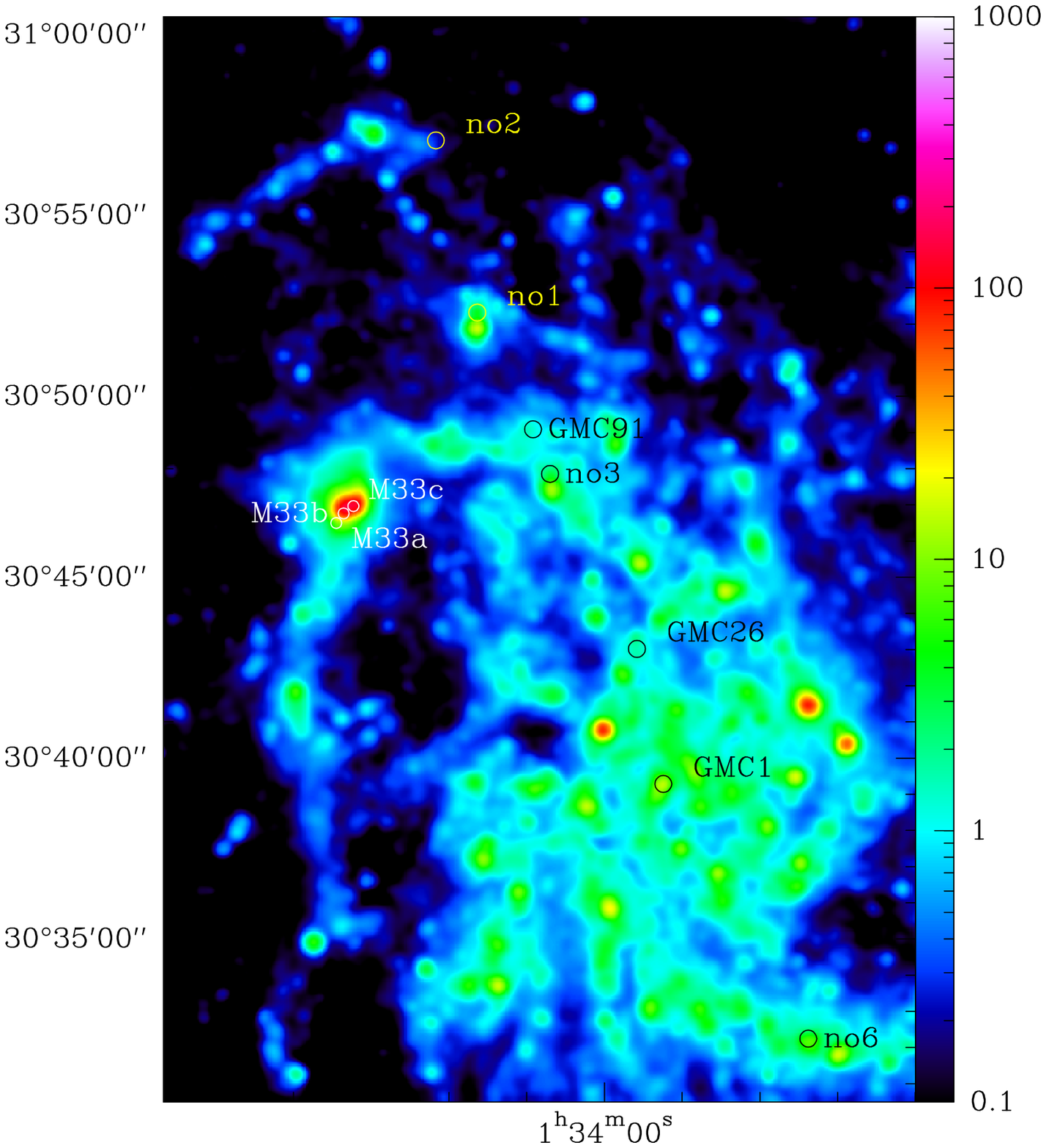}
	\caption{Observed positions on a Spitzer 24$\mu$m map of M~33.  The circle size indicates the beamwidth of the observations.  Units of the 24$\mu$m emission are MJy/sr and the color scale is the same for Figures 1--3 in order to facilitate comparison.  The positions and source names published in \citet{Buchbender13} are shown as well.}
	\label{m33_24mic} 
\end{figure}

We have selected 3 positions in M~33 in/near the giant HII region NGC 604, clouds B8 B9 and B11 identified by \citet{Leroy06} in IC10, and Hubble V in NGC6822 in which HCN was not detected (and HCO$^+$ not observed by \citet{Gratier10b}.  Figures \ref{ic10_24mic},  \ref{n6822_24mic}, and  \ref{m33_24mic} show the positions and beamsizes of the observations on respectively IC~10, NGC~6822, and M~33 as seen at 24$\mu$m with Spitzer.  For M~33, the positions observed in HCN/HCO$^+$ by \citet{Buchbender13} are also shown.  We choose to show the 24$\mu$m emission because it is a high angular resolution tracer of star formation \citep[e.g. ][]{Calzetti2010}.
Table 1 gives the sources, assumed distances, metallicities and  positions of the new observations.

\begin{table}
\begin{center}
\begin{tabular}{lllll}
Source & distance & metallicity & RA & Dec\\
& (kpc) &($Z_{sol}$) & (J2000) & (J2000)   \\
\hline
M~33a & 840 & 0.5 & 01:34:34.5 & 30:46:30  \\
M~33b & 840 & 0.5 & 01:34:33.56 &30:46:46  \\
M~33c & 840 & 0.5 & 01:34:32.3 & 30:46:58  \\
IC10b8 & 950 & 0.35 & 0:20:21.6 & 59:17:09  \\
IC10b9 & 950 & 0.35 & 0:20:22.9 & 59:21:18  \\
IC10b11 & 950 & 0.35 & 0:20:27.7 & 59:17 \\
N6822HubV & 490 & 0.3 &19:44:52.80& -14:43:11  \\
\end{tabular}
\caption[]{Sources and positions observed in M~33, IC~10, and NGC~6822.  Distances are from \citet{Galleti04}  for M~33, \citet{Hunter01} for IC~10, \citet{Mateo98} for NGC~6822. Metallicities are from \citet{Magrini09} for M~33,  \citet{Magrini09b} for IC10, \citet{Skillman89} for NGC~6822.  Positions are chosen from our data \citep{Gratier10,Druard14,Gratier10b} for M~33 and NGC~6822 and \citet{Leroy06} for IC~10. }
\end{center}
\end{table}

\section{Observations and data reduction}

\subsection{IRAM 30meter observations}

The IRAM 30meter telescope \footnote{Based on observations carried out with the IRAM 30m Telescope. IRAM is supported by INSU/CNRS (France), MPG (Germany) and IGN (Spain)}
 was used in Nov/Dec 2014 and May 2015 to observe 3 positions in IC~10, the Hubble V HII region in NGC~6822, and a position in M~33 (M~33a).  The coordinates are given in Table 1.
The E090 and E230 dual polarization sideband-separating EMIR receivers \citep{Carter2012} were used to observe the lines listed in Table 2.  The very broad FTS backends were used, providing 195kHz resolution (channel separation) over up to 16GHz.  The data were taken using the wobbler, with a throw of typically 2 arcminutes, making sure not to risk having emission in the reference position.  Telescope focus was checked 
at the beginning of each day and pointing was checked every 1.5 -- 2 hours. 
A line source, generally DR21 which is well-known to our group, was observed each day to check the tuning.
The beamsize is about 28$''$ at 90GHz, 22$''$ for $^{13}$CO(1--0) and 11$''$ at the frequencies we have used in the 1mm band.

System temperatures varied from 75--120~K for the lines near 90GHz to 100-160 ~K at 110GHz and 200~K at 115GHz and $\sim$300~K in the 1mm band.  Spectra were individually inspected and a small number of spectra identified as being of poor quality were dropped.  The spectra for each line and source were then averaged and a continuum level (i.e. a zero-order baseline) subtracted if necessary.  All spectra are presented on the main beam temperature scale.  The forward efficiency of the telescope is 0.95 in the 3mm band and 0.92 in the 1mm band (218--230GHz).  Conversion to T$_{mb}$ was done using main beam efficiencies of 0.59 for $^{12}$CO(2--1), 0.61 for $^{13}$CO(2--1) and C$^{18}$O(2--1), 0.78 for $^{12}$CO(1--0) and $^{13}$CO(1--0) and C$^{18}$O(1--0), 0.80 for CS(2--1), and 0.81 for HCN, HCO$^+$, HNC, and C$_2$H.  All data processing was done within the GILDAS environment \footnote{http://www.iram.fr/IRAMFR/GILDAS/}.

\begin{table}
\begin{center}
\begin{tabular}{ll}
Line & Frequency  (GHz) \\
\hline
HCN($J=1-0$) & 88.6318475 \\
HCO$^+$($J=1-0$) & 89.18853 \\
HNC($J=1-0$) & 90.66356 \\
CS($J=2-1$) & 97.98095 \\
N$_2$H$^+$($J=1-0$) & 93.1734 \\
C$_2$H($N=1-0$) & 87.316925$^a$ \\
$^{13}$CO($J=1-0$) & 110.20137 \\
C$^{18}$O($J=1-0$) & 109.78218 \\
$^{13}$CO($J=2-1$) & 220.398686 \\
C$^{18}$O($J=2-1$) & 219.56036 \\
\end{tabular}
\caption[]{ HCN has two satellite (hyperfine) lines at relative velocities of $-7.1$ and 4.8 $\kms$ and relative LTE intensities of 0.2 and 0.6 respectively. These lines are generally unresolved in all but the closest extragalactic sources.  HCO$^+$, HNC, and N$_2$H$^+$ have unresolved hyperfine structure.  $^a$ C$_2$H has a total of 6 lines 87.284156, 87.316925(strongest), 87.328624 ($J=3/2 \rightarrow 1/2$)
         87.402004, 87.407165, 87.446512 ($J=1/2 \rightarrow 1/2$). }
\end{center}
\end{table}

\begin{figure}
	\centering
	\includegraphics[width=\hsize{}]{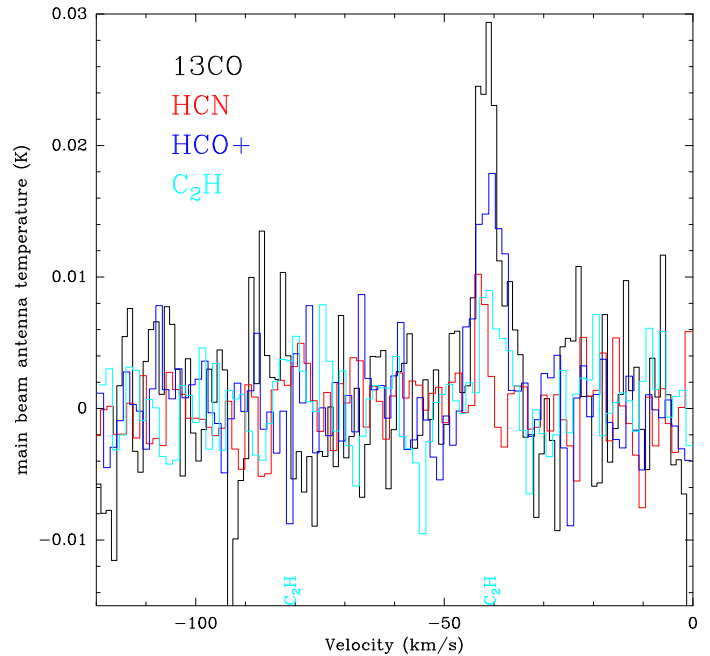}
	\caption{$^{13}$CO and dense gas tracers as observed in the Hubble V HII region in NGC~6822.  Color coding is indicated in the panel.}
	\label{HubV} 
\end{figure}

\begin{figure}
	\centering
	\includegraphics[width=\hsize{}]{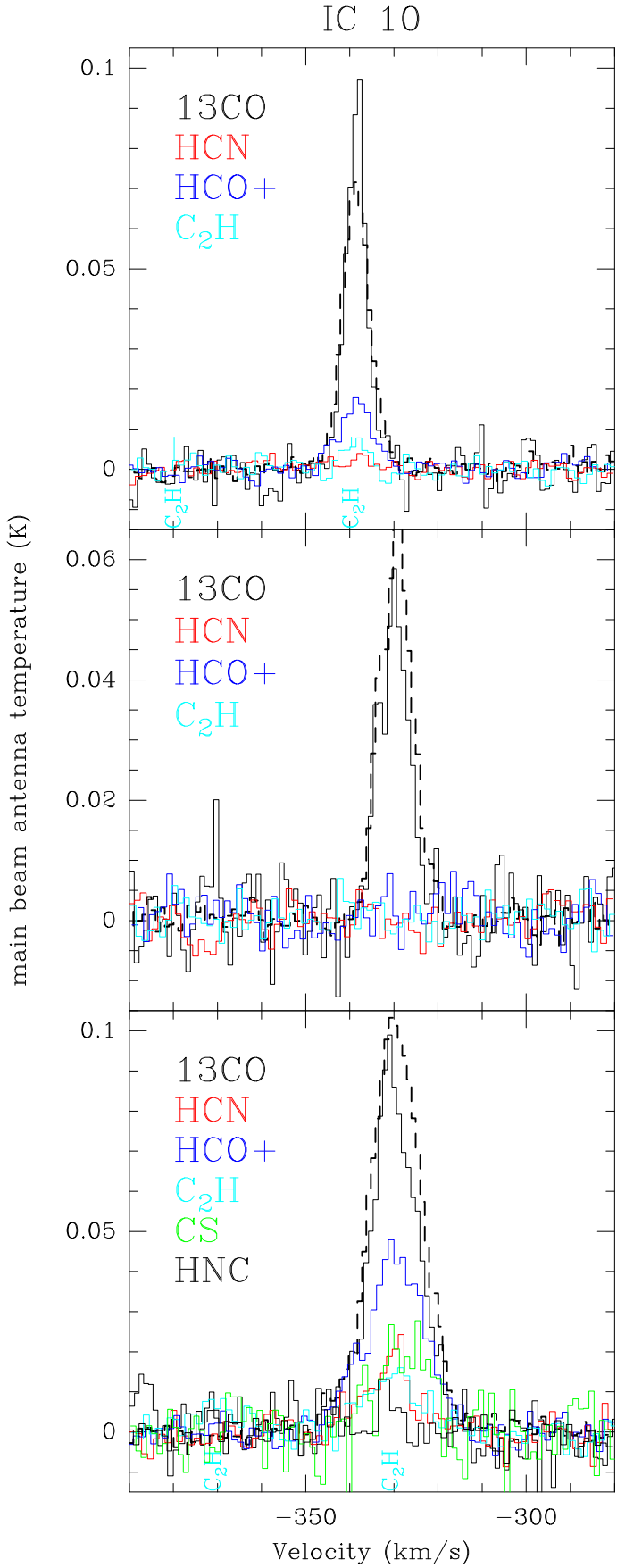}
	\caption{$^{13}$CO, $^{12}$CO and dense gas tracers as observed in the b8, b9, and b11 regions in IC~10.  Color coding is indicated in the panel.  Note that additional lines were detected in IC~10 b11.  The $^{12}$CO line scale (dashed) is divided by 10.}
	\label{IC10} 
\end{figure}

\begin{figure}
	\centering
	\includegraphics[width=\hsize{}]{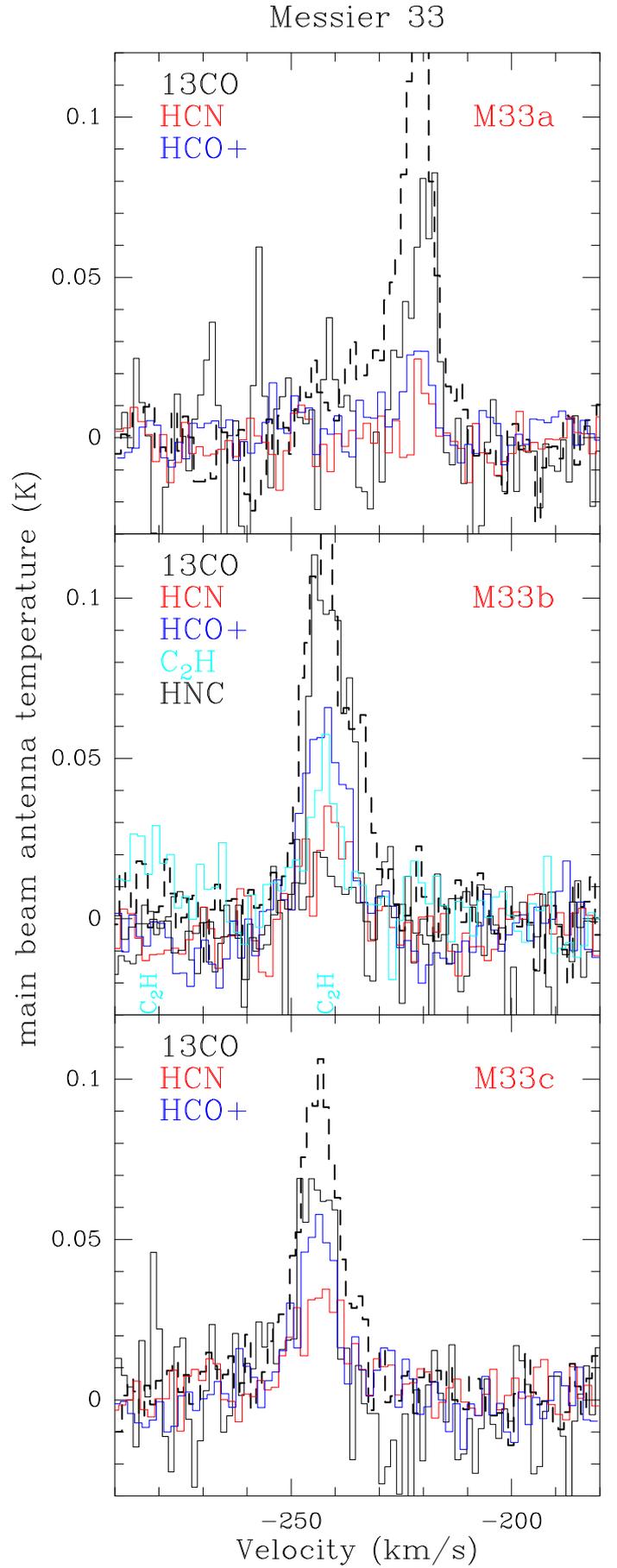}
	\caption{$^{13}$CO, $^{12}$CO and dense gas tracers as observed in the M~33a, M~33b, and M~33c regions in M~33, all near the giant HII region NGC~604.  Color coding is indicated in the panel.  Note that additional lines were detected in M~33b.  The $^{12}$CO line scale (dashed) is divided by 10.}
	\label{M33} 
\end{figure}

\begin{figure}
	\centering
	\includegraphics[width=\hsize{}]{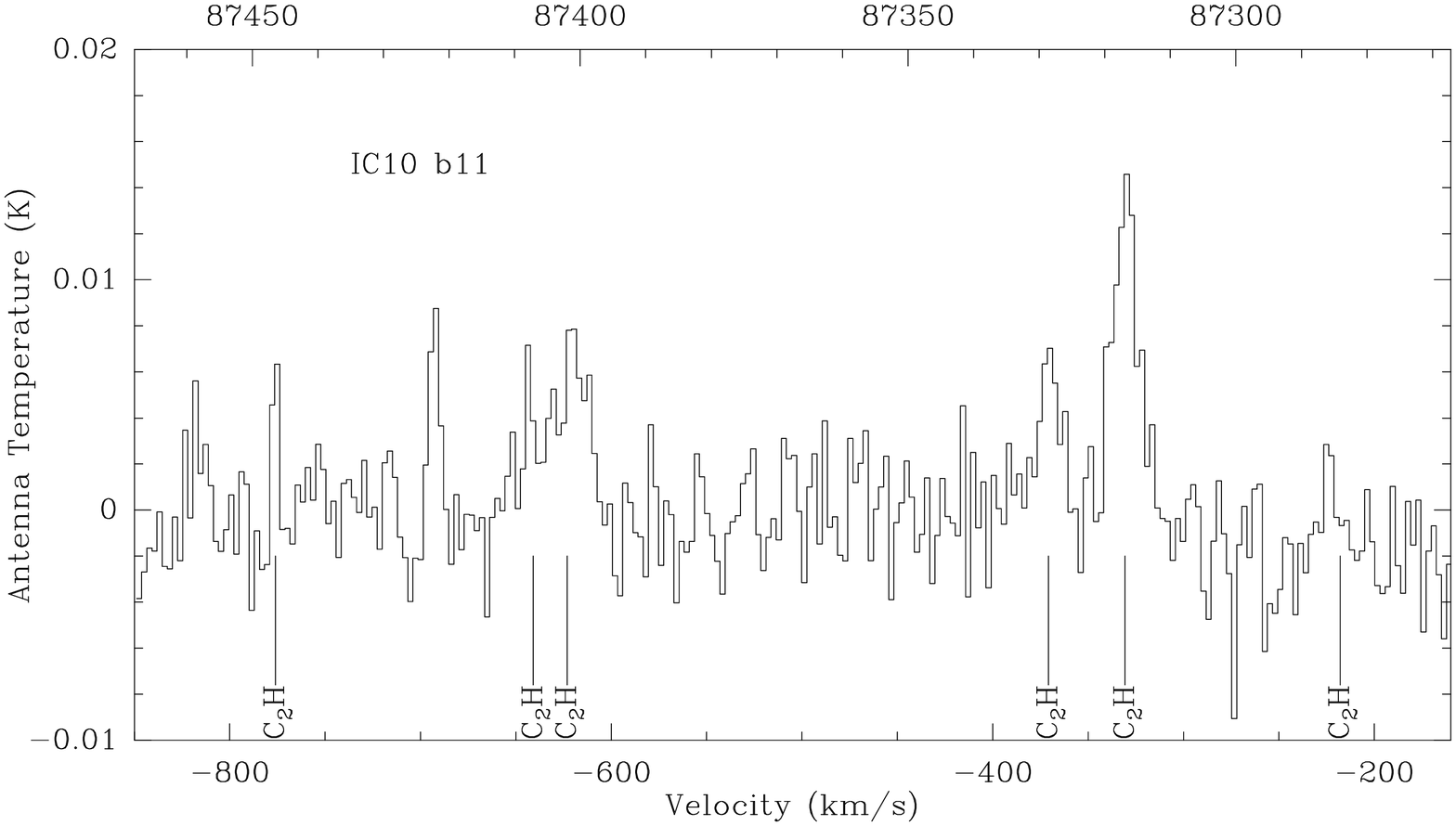}
	\caption{C$_2$H spectrum observed in source b11 in the galaxy IC10.  Upper scale gives the rest frequency in MHz.  The 6 transitions are marked.  The flux ratios are compatible with LTE.}
	\label{ic10_c2h} 
\end{figure}

\subsection{Nobeyama Radio Observatory 45meter observations}

The Nobeyama 45 m observations took place from May 20th to June 1st, 2015 toward the GMCs in M~33 near NGC~604.
The data were taken in the position-switching mode. The TZ receiver, which is a
dual polarization sideband-separating SIS receiver, was used in combination with
the Fast Fourier Transform Spectrometer SAM45. To cover the $^{12}$CO (1--0),
$^{13}$CO(1--0), C$^{18}$O(1--0), HCO$^+$(1--0), HCN(1--0), C$_2$H(1--0), HNC(1--0), SO(2,3--1,2), HC$_3$N
(10--9) frequencies, we used three frequency settings with a bandwidth of 8 GHz (1 GHz x
8) and a frequency resolution of 244.14 kHz in the dual polarization mode. Pointing was
checked by observing the IRC+30021 SiO maser emission every hour, and was shown to
be accurate within 3 arcseconds. The main-beam efficiency was 38.8 - 50.0\%.  The
system noise temperature range was 140 -- 500 K.  The beamsize of the NRO 45m telescope 
is about 19$''$ at 90 GHz and 15$''$ at 115 GHz.

We used the 4.3.1 version of the Common Astronomy Software Application (CASA)
package \citep{McMullin07} to reduce the data obtained with the Nobeyama 45
m. The CASA package is developed for the Atacama Large Millimeter/submillimeter
Array (ALMA), and is also available for reducing data obtained by a single-dish
telescope such as the Nobeyama 45 m telescope. Here, we describe
detailed steps of the data reduction process according to \citet{Shimajiri2015}. 
In the first step, edge channels of each correlator band were flagged 
using the task $sdflag2old$, since the sensitivity of the edge channels drops. The
SAM45 spectrometer of the Nobeyama 45 m telescope provides 4096 frequency
channels, so that the data for the channel numbers from 1 to 300 and from 3796 to
4096 were flagged. In the second step, we determined the baseline of each
spectrum using the task $sdbaselineold$. In the third step, we merged all the
averaged spectra into one file using the task $sdcoadd$. In the final step, we applied
the baseline correction again using the task $sdbaselineold$.



\begin{table*}
\begin{center}
\begin{tabular}{lllllllll}
Source & I$_{CO}$ & I$_{^{13}CO}$ & I$_{C^{18}O}$ &I$_{HCO^+}$  & I$_{HCN}$ &  I$_{HNC}$ &I$_{CCH}$&I$_{CS}$  \\
\hline
M~33a$^a$ & & $0.64\pm .07$ & $0.04\pm.065$& $0.20\pm .05$ &$0.12\pm .05$ &$\la 0.06$ &$0.09\pm .05$ &  \\
M~33a$^b$ & $15.7\pm .6$ &$0.58\pm .10$ &$0.21\pm .12$ &$0.212\pm .037$ &$0.083\pm .039$ &$0.031\pm .036$ &$0.117\pm .048$ & \\
M~33b$^b$ & $15.5\pm .5$ &$1.16\pm .12$ &$\la 0.11$ &$0.806\pm .055$ &$0.339\pm .048$ &$0.208\pm .046$ &$0.434\pm .068$  & \\
M~33c$^b$ & $12.1\pm .4$ &$0.73\pm .08$ &$\la 0.09$ &$0.659\pm .046$ &$0.496\pm .046$ &$\la 0.051$ &$0.075\pm .060$  \\
IC10b8$^a$ & $5.54\pm .08$ & $0.55\pm .023$ & $\la 0.014$ & $0.157\pm .009$ & $0.032\pm .010$ & $0.025\pm .009$ & $0.031\pm .009$ & $\la 0.04$  \\
IC10b9$^a$ &  $5.93\pm .08$ &  $0.46\pm .03$ &  $\la 0.02$ &  $0.04\pm .02$ &  $\la 0.015$&  $\la 0.016$ &  $\la 0.014$ &  $0.06\pm .03$  \\
IC10b11$^a$ &  $14.9\pm .1$ & $1.12 \pm .04$ & $0.11 \pm .036$  & $0.718 \pm .023$  & $0.245 \pm .025$  & $0.088 \pm .027$  & $0.21 \pm .03$  & $0.26 \pm .05$  \\
N6822HubV$^a$ & $2.0\pm 0.12^c$ & $0.16\pm .02$ & $\la 0.019$ & $0.115\pm .012$ & $0.025\pm .011$ & $\la 0.013$ & $0.049\pm .012$   \\
\hline
\end{tabular}
\caption[]{ New observations presented in this work. All fluxes are on the main beam temperature scale. $^a$ indicates observatons with the IRAM 30meter antenna and $^b$ with the NRO 45meter telescope.  $^c$ taken from \citet{Gratier10b}.  The other lines within the bandpasses such as SO, HC$_3$N, or N$_2$H$^+$, were not detected.  C$^{18}$O was not detected either but was included in this table to directly compare with $^{13}$CO.  Blank parts of the table indicate that the line was not observed.  Limits are given as $\la$ the 1$\sigma$ noise level to distinguish between a tentative line (e.g. HCN in N6822HubV, $0.25 \pm .011$ Kkm/s) and something where there is no sign of a line (typically intensity below 1 sigma or negative).} 
\end{center}
\end{table*}


\section{Spectra and line ratios}

The line intensities observed as part of the new observations are presented in Table 3.  In all cases, the C$^{18}$O lines were not detected with noise levels that place the C$^{18}$O at typically 10\% or less of the $^{13}$CO line intensity.  The weakness of the C$^{18}$O relative to the more abundant isotopes is even more extreme in the Magellanic Clouds \citep{Chin97,Chin98}, where $^{13}$CO/C$^{18}$O line ratios are $\sim 40$, including some lower limits.  In contrast, the line survey of two positions in M~51 carried out recently by \citet{Watanabe14} showed $^{13}$CO/C$^{18}$O line ratios of 4 although the $^{12}$CO/$^{13}$CO line ratios are $\sim 10$, similar to those of the other galaxies.  The weak C$^{18}$O emission found in these galaxies, including M~33, indicates that low $^{18}$O abundance is a characteristic of subsolar metallicity environments although selective photodissociation of C$^{18}$O likely further decreases the C$^{18}$O emission \citep{Shimajiri2014}.

Figures 4 -- 6 present the spectra of the main (detected) lines of interest -- $^{13}$CO(1--0), HCN(1--0), HCO$^+$(1--0), 
C$_2$H, and  $^{12}$CO(1--0).  It is immediately apparent that the HCN lines are weak relative to HCO$^+$.  This was already seen in the observations of the Magellanic Clouds presented by \citet{Chin97} and  \citet{Chin98}.  It was also noted for M~33 by \citet{Buchbender13}, as compared to M~31 \citep{Brouillet05} where the HCN and HCO$^+$ intensities are similar.   \citet{Watanabe14} find stronger HCN than HCO$^+$ emission near the nucleus of M~51.  \citet{Bigiel16} found generally stronger HCN emission compared to HCO$^+$ in the inner disk of M~51.   
The metallicity of M~51 is solar or somewhat higher and decreases with distance from the galactic center \citep{Bresolin2004}.  
Several clouds in the outskirts of M~51 were observed by \citet{Chen2016} 
who found similar HCN and HCO$^+$ fluxes (slightly stronger HCO$^+$).  All these works found that the HNC emission from M~51 was 20--50\% of the HCN strength.  This HNC/HCN ratio is consistent with our data and those presented by \citet{Chin97} and  \citet{Chin98} but higher than found by \citet{Buchbender13} in their stacked spectrum.

The spectra presented by \citet{Watanabe14} can be straightforwardly compared with our own.  They detect the N$_2$H$^+$  and HNCO lines at a level stronger than the C$_2$H lines and 5--10\% of the $^{13}$CO(1--0) line strength.  These lines are not detected in the low-metallicity systems.  The CN lines are strong in M~51, 10--15\% of the $^{13}$CO(1--0) line strength.  The $^{13}$CO/$^{12}$CO line ratio is nearly constant among the sources, almost irrespective of metallicity, so we choose these lines as a reference to compare relative line strengths.  Neither we nor \citet{Buchbender13} detect CN, even in stacked spectra.
However, since we observed CN simultaneously with $^{12}$CO(1--0), a strong line, the limits are of order 10\% of the $^{13}$CO(1--0) intensity, and less constraining in some cases.
The limits to the CN emission show, for the \citet{Buchbender13} sources and IC10 (the other upper limits in our data are not constraining), that it is weaker than HCN and C$_2$H but consistent with the observations of the LMC by \citet{Chin97}.

In addition to C$^{18}$O, the weak and/or non-detected lines are all of Nitrogen bearing molecules.  Not only is the  CO luminosity per H$_2$ mass low in subsolar metallicity systems but the Nitrogen-bearing molecules 
are weak relative to CO.  The
 C$_2$H lines on the other hand are not weak relative to CO.  Figure \ref{ic10_c2h} shows the clearest of our C$_2$H spectra, including most or all of the 6 components.  While the S/N ratio is low for the weaker components, the fluxes are consistent with the ratios expected at Local Thermodynamic Equilibrium (LTE). 

\citet{Tercero2010} observed Orion KL with the IRAM 30meter telescope.  As can be seen from their Fig. 2, the HCN and H$^{13}$CN lines are much stronger than the HCO$^+$ and H$^{13}$CO$^+$ lines respectively.  These lines are just an example as generally, in this gas exposed an extremely high radiation field, the line ratios are different both from M~51 and lower metallicity sources.
\citet{Nishimura2016} observed position IC10b11 with the NRO 45m and obtain similar detections and non-detections.



\section{Source of the variations in line ratios}

HCN and HCO$^+$ have similar dipole moments and are both linear molecules, with a substitution 
of O$^+$ for N in the latter.  It is thus reasonable to expect a link between the N/O abundance ratio 
and the abundance ratio of the two molecules.  

For the range in metallicities of our sources, the N/O abundance ratio is roughly proportional to the Oxygen abundance (O/H) although the N/O ratio appears roughly constant for lower metallicities \citep[see Figures 3--4 and 5 in ][ respectively]{Pilyugin03,vanZee06}.  This is attributed to a change in the origin of Nitrogen from primary at low metallicity to secondary at higher metallicity.

\begin{figure}
	\centering
	\includegraphics[width=\hsize{}]{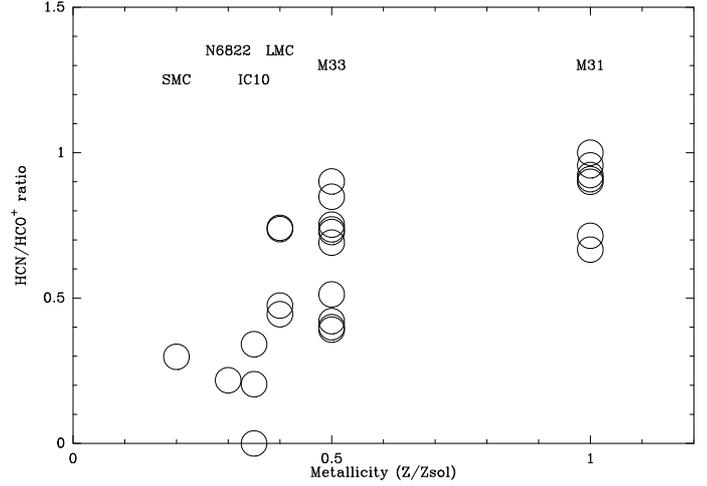}
	\caption{Variation of the HCN/HCO$^+$ ratio with metallicity.  References are \citet{Brouillet05, Chin97, Chin98} for M~31 and the Magellanic Clouds, \citet{Buchbender13} and the present work for M~33, and this work for IC~10 and NGC~6822.   Typical uncertainties for individual points are 0.2 dex for the metallicity and 0.3 in the HCN/HCO$^+$ ratio.}
	\label{hcnop_z} 
\end{figure}

Figure \ref{hcnop_z} shows the variation of the HCN/HCO$^+$ line intensity ratio.  Both of these lines are optically thick in dense regions of the Galaxy but at these metallicities and scales (10-100 pc), the lines are expected to be optically thin, with the possible exception of M~31. 
In Fig. \ref{hcnop_z}, each galaxy has been assumed to have the same metallicity for all positions.  M~33 and M~31 only have weak metallicity gradients \citep{Magrini09,Sanders2012} and the others have no known gradient although there may be metallicity inhomogeneities.
 An increase in the HCN/HCO$^+$ line intensity ratio from roughly 1/4 at a metallicity of 0.3 solar to about 0.9 in M~31 follows this trend in N/O abundance.  At these scales, we have no reason to invoke other sources than the N/O abundance ratio to explain the observed trend in the HCN/HCO$^+$ line intensity ratio with metallicity.
The N/O abundance ratio was also proposed by \citet{Anderson2014} as a possible explanation of the low HCN/HCO$^+$ ratio in the LMC.

\section{Link with star formation}

\begin{figure}
	\centering
	\includegraphics[width=\hsize{}]{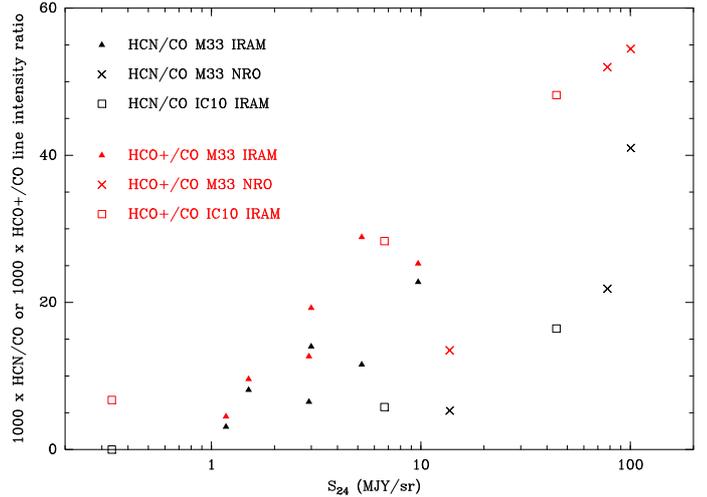}
	\caption{Link between dense gas fraction and star formation rate.  The HCN/CO and HCO$^+$/CO ratios are used to trace the fraction of dense molecular gas and the 24$\mu$m intensity is used as a proxy for the SFR.  HCN/CO is in black and HCO$^+$/CO is in red.  Observations are from \citet{Buchbender13} and the present work. Typical uncertainties are 20\% for HCO$^+$/CO and 25\% for the HCN/CO ratio.}
	\label{hcn_ir} 
\end{figure}

Figure \ref{hcn_ir} shows the link between the dense gas fraction, as traced by the intensity ratios of HCN and HCO$^+$ to CO, and the SFR as traced by the 24$\mu$m emission.  When the 24$\mu$m emission is strong, both the HCN/CO and HCO$^+$/CO ratios are high and the relation appears monotonic.  However, the HCO$^+$/CO ratio shows less scatter between the galaxies (M~33 and IC10) than the HCN/CO line ratio, presumably due to the lesser effect of metallicity variations which cause the HCN/CO ratios of IC10 to be lower than in M~33.
Therefore, at the 100pc scale, it appears that HCO$^+$ may be the preferred tracer of dense gas in low-metallicity objects.  

\citet{Buchbenderthesis} observed HCN, HCO$^+$, and CO in cloud no3 in M~33 at a resolution of about 20pc.  From his figure 11.3, clear differences at this scale are seen between the distribution of the two lines.  His figure 11.5 shows a comparison of the HCN and HCO$^+$ maps with various star formation tracers.  

\begin{figure}
	\centering
	\includegraphics[width=\hsize{}]{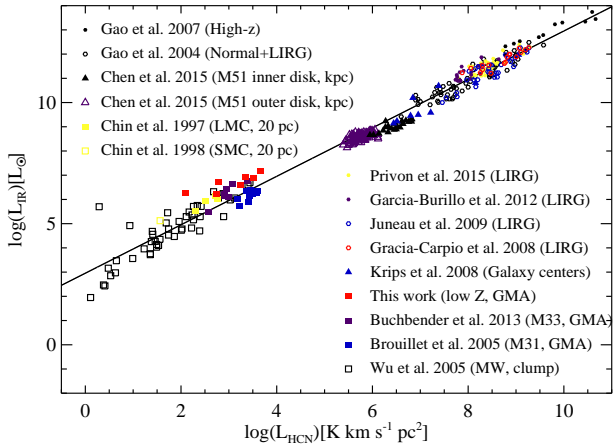}
	\includegraphics[width=\hsize{}]{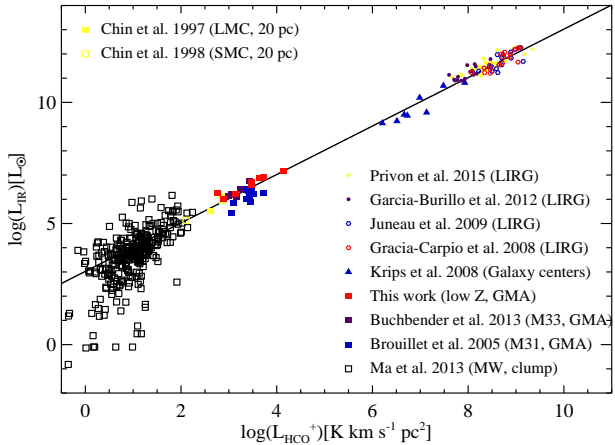}
	\caption{Link between IR luminosity and HCN (upper panel) and HCO$^+$ (lower panel) luminosity.  
	The lines show a linear relation between the two quantities.   The fits are $log(L_{IR}) = 2.95424 + log(L_{HCN})$ 
	and $log(L_{IR}) = 3.04258 + log(L_{HCO^+})$ and are from respectively \citet{Gao04} and 
	\citet{Gracia-Carpio08}. 
	The symbols indicate the reference and the scale of the observation.  }
	\label{hcnhcop-IR} 
\end{figure}

The observations presented here are at a scale of 100pc (range of 70--140pc), intermediate between the subpc-scale Galactic data and the typically kpc or whole-galaxy extragalactic measurements.  Figure~\ref{hcnhcop-IR} shows how the HCN and HCO$^+$ luminosities compare with Galactic and extragalactic observations at various scales.
Points above the line can be considered to have weak line emission and those below to have strong line emission.
The trend is basically a linear relation between the emission of the dense gas tracers and IR luminosity, presumably reflecting a proportionality between the mass of dense gas and the star formation rate.  However, and this is why we have chosen to separate galactic centers and inner disks from outer regions, the physical conditions in large galaxies are such that stellar surface densities and molecular fractions are systematically higher in the inner disk as compared to external regions \citep{Chen2015,Usero2015}.  Furthermore, spirals show radial metallicity gradients. 

\begin{figure}
	\centering
	\includegraphics[width=\hsize{}]{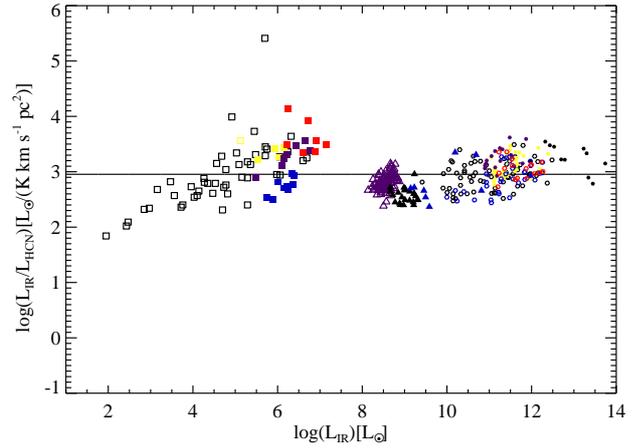}
	\includegraphics[width=\hsize{}]{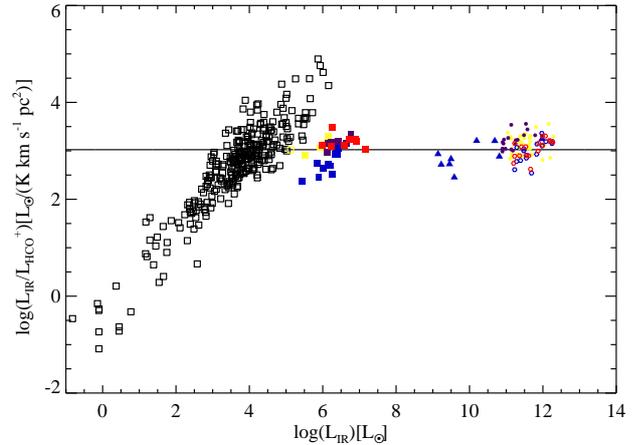}
	\caption{Ratio between IR and HCN (upper panel) and HCO$^+$ (upper panel) luminosities as a function of IR luminosity.  The goal here is to see finer structure than in the previous figure as the vertical scale has been greatly expanded.  References are as given in Fig.~\ref{hcnhcop-IR}. }
	\label{hcnhcop-IR_ratio} 
\end{figure}

Looking at the data points above and below the line showing HCN-IR proportionality, it is immediately clear that HCN is relatively stronger in the inner regions 
\citep{Chen2015,Krips2008} and weaker in low-metallicity galaxies \citep[points from ][ and this work]{Chin97,Chin98,Buchbender13}.  Turning to the HCO$^+$ emission in the lower panel, the linear relation appears to have less systematic scatter in that the low-Z points are not off the global fit and inner disk points are closer to it.  However, fewer extragalactic HCO$^+$ data are available.
Figure~\ref{hcnhcop-IR_ratio} expands the vertical scale by plotting the ratio of the IR-to-HCN (or HCO$^+$) luminosity as a function of the IR luminosity.  Here it becomes apparent that at the small scales observed by \citet{Ma2013} and \citet{Wu2005}, the IR emission increases more rapidly than the HCN or HCO$^+$.  This could be due to either an increase in the line optical depth, causing saturation for the more massive clumps, or, as suggested by \citet{Wu2010}, that the IR luminosity of low-mass star forming regions decreases more quickly due to an incomplete sampling of the IMF such that few (if any) massive stars are present.  

One would expect high redshift objects to have generally subsolar metallicities and, due to their necessarily young ages, low N/O abundance ratios (much of the Nitrogen production occurs after a substantial time delay).  
The observations of high-redshift objects, necessarily very bright, by \citet{Gao07} seem to go in this direction as the HCN emission per unit IR luminosity is weak (or equivalently the IR emission is strong).  
The HCN/IR ratio is about a factor of 2 weaker compared to that established in local galaxies, yet 
this value is similar to what we found here in nearby low-Z galaxy systems. Nevertheless, the HCN/CO 
ratios of the low-Z systems and high-redshift AGNs/ULIRGs are at the two extreme ends, the smallest 
HCN/CO values are found in low-Z whereas the highest are in high redshift AGNs/ULIRGs, similar to that of local 
ULIRGs despite their higher HCN/IR ratio. It is still unclear whether low N/O abundance ratios could explain 
all the differences. Therefore, the low HCN/IR ratio at high redshift might suggest a higher star formation rate 
per unit dense molecular gas in high-redshift AGNs/ULIRGs, although not necessarily so in nearby low-Z 
systems, compared to those of local normal star-forming galaxies.  HCO$^+$ measurements would greatly help to 
determine the appropriate interpretation.


\section{Summary and Conclusions}

 The low-metallicity Local Group galaxies NGC~6822, IC~10, and M~33 have been observed in a number of molecular lines, including the dense gas tracers HCN(1--0) and HCO$^+$(1--0), at a scale of 70--140 pc.
In addition to $^{12,13}$CO, HCO$^+$, HCN, HNC, C$_2$H, and CS(2--1) have been detected in one or more objects.

The HCN and HNC lines are weak with respect to the IR or $^{12,13}$CO emission but the HCO$^+$ follows the trends observed in solar metallicity galaxies.  C$^{18}$O, N$_2$H$^+$, CN, and HNCO were not detected although in M~51 \citep{Watanabe14} these lines are stronger than C$_2$H, which was well detected.
N/O varies as O/H at the metallicities explored here.  An accordingly low Nitrogen abundance could explain the weakness of the Nitrogen-bearing species in our observations and more generally in the observations of low metallicity galaxies.

The SFR, as traced by the 24 um emission, increases with the 
dense gas fraction, as traced by the HCN/CO or HCO$^+$/CO line ratios.

The low--Z points fall below (i.e. less HCN) the linear HCN--IR Galactic-extragalactic luminosity correlation but right on the HCO$^+$--IR correlation.  When measuring the dense gas mass of low metallicity objects, HCO$^+$ may be a more robust tracer than HCN.


 



   
\begin{acknowledgements}
JB would like to thank Laura Magrini for very useful discussions about abundances and abundance gradients in galaxies.  We would also like to thank Belen Tercero for the the data enabling us to calculate the line ratios of interest in Orion KL.  The authors would like to acknowledge support from the French Agence Nationale de Recherche (ANR) grant ANR-11-BS56-010 for the STARFICH project ( www.obs.u-bordeaux1.fr/webformationetoiles/StarFichE ) and the European Research Council under the European Union's Seventh Framework Programme (ERC Advanced Grant Agreement no. 291294 --  `ORISTARS').  
YG acknowledges support from the National Natural Science Foundation of China (NSFC grants 11390373 and 11420101002), the Strategic Priority Research Program "The Emergence of Cosmological Structures" (grant XDB09000000) and Key Research Program of Frontier Sciences of the Chinese Academy of Sciences.
\end{acknowledgements}

\nocite{Ma2013}
\nocite{Juneau2009}
\nocite{Privon2015}

\bibliographystyle{../aa}
\bibliography{../jb}

\end{document}